\documentclass[12pt,letterpaper]{article}
\pdfoutput=1
\usepackage{scicite}
\usepackage{times}
\usepackage{upgreek}

\newcommand{\lsim}{\,\rlap{\raise 0.35ex\hbox{$<$}}{\lower 0.7ex\hbox{$\sim$}}\,}
\newcommand{\gsim}{\,\rlap{\raise 0.35ex\hbox{$>$}}{\lower 0.7ex\hbox{$\sim$}}\,}

\usepackage{graphicx}
\usepackage[usenames]{color}
\def \aap{Astron.~Astrophys.}
\def \aj{Astron.~J.}
\def \al{Astron.~Lett.}
\def \apj{Astrophys.~J.}
\def \apjl{Astrophys.~J.}
\def \apjs{Astrophys.~J.~Suppl.}
\def \cup{Cambridge~Univ.~Press}
\def \epjd{Euro.~Phys.~J.~D}
\def \jqe{IEEE~J.~Quant.~Electron.}
\def \mnras{Mon.~Not.~R.~Astron.~Soc.}
\def \nat{Nature}
\def \ol{Opt.~Lett.}
\def \prl{Phys.~Rev.~Lett.}
\def \va{Vistas~Astron.}

\newenvironment{sciabstract}{%
\begin{quote} \bf}
{\end{quote}}

\newcounter{lastnote}

\title{Laser Frequency Combs for Astronomical Observations}

\author{Tilo~Steinmetz,$^{1,2}$ Tobias~Wilken,$^{1}$ Constanza~Araujo-Hauck,$^{3}$\\
  Ronald~Holzwarth,$^{1,2}$ Theodor~W.~H\"ansch,$^{1}$ Luca~Pasquini,$^{3}$\\
  Antonio~Manescau,$^{3}$ Sandro~D'Odorico,$^{3}$ Michael~T.~Murphy,$^{4}$\\
  Thomas~Kentischer,$^{5}$ Wolfgang Schmidt,$^{5}$ and Thomas~Udem$^{1\ast}$\\
\\
\normalsize{$^{1}$Max-Planck-Institut f\"ur Quantenoptik, Hans-Kopfermann-Strasse 1, D-85748 Garching, Germany}\\
\normalsize{$^{2}$Menlo Systems GmbH, Am Klopferspitz 19, D-82152 Martinsried, Germany}\\
\normalsize{$^{3}$European Southern Observatory, Karl-Schwarzschild-Strasse 3, D-85748 Garching, Germany}\\
\normalsize{$^{4}$Centre for Astrophysics and Supercomputing, Swinburne University of Technology,}\\
\normalsize{Mail H39, PO Box 218, Victoria 3122, Australia}\\
\normalsize{$^{5}$Kiepenheuer-Institut f\"ur Sonnenphysik, Sch\"oneckstr. 6, D-79104 Freiburg, Germany}\\
\\
\normalsize{$^\ast$To whom correspondence should be addressed. E-mail:  thu@mpq.mpg.de}
}

\date{}

\begin{document} 


\maketitle


\begin{sciabstract}
  A direct measurement of the universe's expansion history could be
  made by observing in real time the evolution of the cosmological
  redshift of distant objects. However, this would require
  measurements of Doppler velocity drifts of
  $\sim$1\,centimeter per second per year, and astronomical spectrographs have
  not yet been calibrated to this tolerance. We demonstrate the
  first use of a laser frequency comb for wavelength calibration of an
  astronomical telescope. Even with a simple analysis, absolute
  calibration is achieved with an equivalent Doppler precision of
  $\sim$9\,meter per second at $\sim$1.5\,micrometers---beyond
  state-of-the-art accuracy. We show that tracking complex,
  time-varying systematic effects in the spectrograph and detector
  system is a particular advantage of laser frequency comb
  calibration. This technique promises an effective means for
  modeling and removal of such systematic effects to the accuracy
  required by future experiments to see direct evidence of the
  universe's putative acceleration.
\end{sciabstract}

Recent cosmological observations suggest that the universe's expansion
is accelerating. Several lines of evidence corroborate this, including
results from distant supernovae~\cite{RiessA_98a,PerlmutterS_99a}, the
cosmic microwave background~\cite{SpergelD_03a}, and the clustering of
matter~\cite{PeacockJ_01a,EisensteinD_05a}. However, the current
observations are all essentially geometric in nature, in that they map
out space, its curvature, and its evolution. In contrast, a direct and
dynamical determination of the universe's expansion history is
possible by observing the slow drift of cosmological redshifts, $z$,
that is inevitable in any evolving universe~\cite{SandageA_62a}.  No
particular cosmological model or theory of gravity would be needed to
interpret the results of such an experiment. However, the cosmological
redshift drift is exceedingly small and difficult to measure; for
currently favored models of the universe, with a cosmological constant
parametrizing the acceleration, the redshifts of objects drift by less
than $\sim$1\,cm\,s$^{-1}$\,year$^{-1}$ (depending on their redshifts).

Nevertheless, the suggestion that the so-called Lyman-$\alpha$
"forest" seen in high-redshift quasar spectra is the best target for
this experiment~\cite{LoebA_98a} was recently supported by cosmological
hydrodynamical simulations~\cite{LiskeJ_08a}. The forest of absorption
lines is caused by the Lyman-$\alpha$ transition arising in neutral
hydrogen gas clouds at different redshifts along the quasar
sight-lines. Detailed calculations with simulated quasar spectra show
that the planned 42-m European Extremely Large Telescope (E-ELT),
equipped with the proposed Cosmic Dynamics Experiment (CODEX)
spectrograph~\cite{PasquiniL_06a}, could detect the redshift drift
convincingly with 4000\,hrs of observing time over a $\sim$20-year
period~\cite{LiskeJ_08a}. Therefore, as the observation is feasible (in
principle), overcoming the many other practical challenges in such a
measurement is imperative. Important astrophysical and technical
requirements have been considered in detail, and most are not difficult
to surmount~\cite{LiskeJ_08a,materials}. One (but not the only)
extremely important requirement is that the astronomical spectrographs
involved must have their wavelength scales calibrated accurately
enough to record $\sim$1\,cm\,s$^{-1}$ velocity shifts ($\sim$25-kHz
frequency shifts) in the optical range. Moreover, this accuracy must
be repeatable over $\sim$20-year timescales.

Although the redshift drift experiment requires demanding
precision and repeatability, precisely calibrated astronomical
spectrographs have several other important applications. For example,
Jupiter- and Neptune-mass extrasolar planets have been discovered by
the reflex Doppler motion of their host stars~\cite{MayorM_95a,MarcyG_96a,LovisC_06a}, but detecting Earth-mass
planets around solar-mass stars will require $\sim$5\,cm\,s$^{-1}$
precision maintained over several-year time scales~\cite{LovisC_06b}.
Another example is the search for shifts in narrow quasar absorption
lines caused by cosmological variations in the fundamental constants
of nature~\cite{BahcallJ_65a,WebbJ_99a,ThompsonR_75a}. Recent
measurements~\cite{MurphyM_03a,ChandH_04a,LevshakovS_06b,ReinholdE_06a} achieve
precisions of $\sim$20\,m\,s$^{-1}$, but the possibility of hidden
systematic effects, and the increased photon-collecting power of
future ELTs, warrant much more precise and accurate calibration over
the widest possible wavelength range.

Laser frequency combs (LFCs) offer a solution because they provide an
absolute, repeatable wavelength scale defined by a series of laser
modes equally spaced across the spectrum. The train of femtosecond
pulses from a mode-locked laser occurs at the pulse repetition rate,
$f_{\rm rep}$, governed by the adjustable laser cavity length. In the
frequency domain, this yields a spectrum, $f_n = f_{\rm ceo} + (n\times
f_{\rm rep})$, with modes enumerated by an integer $n\sim10^5$\,to\,$10^6$.
The carrier envelope offset frequency, $f_{\rm ceo} \le f_{\rm rep}$,
accounts for the laser's internal dispersion, which causes the group
and phase velocities of the pulses to differ~\cite{UdemT_02a}. Thanks to the large integer $n$, the optical frequencies $f_n$ are at hundreds of THz whereas both $f_{\rm rep}$ and $f_{\rm ceo}$ are radio frequencies and can be handled with simple
electronics and stabilized by an atomic clock~\cite{UdemT_02a}. Each
mode's absolute frequency is known to a precision limited only by the
accuracy of the clock. Even low-cost, portable atomic clocks provide
$\sim$1\,cm\,s$^{-1}$ (or 3 parts in 10$^{11}$) precision. Because
LFC light power is much higher than required, the calibration
precision possible is therefore limited by the maximum signal-to-noise
ratio (SNR) achievable with the detector. For modern astronomical
charge-coupled devices (CCDs), the maximum SNR in a single exposure is
limited by their dynamic range but is still sufficient to achieve
$\sim$1\,cm\,s$^{-1}$ precision~\cite{MurphyM_07e}. Furthermore, because
LFC calibration is absolute, spectra from different epochs, or even
different telescopes, can be meaningfully compared.

The main challenge in reaching $\sim$1\,cm\,s$^{-1}$ calibration
accuracy will be the measurement and, eventually, mitigation and/or
modeling and removal of systematic effects in astronomical spectrographs
and detectors. For typical high-resolution spectrographs, a
$\sim$1\,cm\,s$^{-1}$ shift corresponds roughly to the physical size
of a silicon atom in the CCD substrate. Only with the statistics of a
very large number of calibration lines can the required sensitivity be
achieved, provided that systematic effects can be controlled
accordingly~\cite{materials}. For example, even in a highly
stabilized, vacuum-sealed spectrograph, small mechanical drifts will
slightly shift the spectrum across the CCD. Although this can easily be
tracked to first order, other effects such as CCD intrapixel
sensitivity variations will be important for higher precision.
Discovering, understanding and eventually modeling and removing these
effects is crucial for the long-term goal of accurate
calibration; tests of LFCs on astronomical telescopes, spectrographs
and detectors are therefore imperative.

We have conducted an astronomical LFC test on the German Vacuum Tower
Telescope~\cite{SchroeterE_85a} (VTT) (Fig.~1). We used a portable
rubidium clock with a modest accuracy of 5 parts in $10^{11}$ (or
1.5\,cm\,s$^{-1}$); much more accurate clocks are available if needed. This
sets the absolute uncertainty on the frequency of any given comb mode.
The VTT can be operated at near-infrared wavelengths, thereby
allowing a relatively simple and reliable fiber-based LFC to be used.
The erbium-doped fiber LFC used had $f_{\rm rep}=250$\,MHz which,
despite the VTT spectrograph having higher resolving power (resolution
of 0.8\,GHz or 1.2\,km\,s$^{-1}$) than most astronomical
spectrographs, is too low for modes to be resolved apart. Filtering
out unwanted modes by using a Fabry-P\'erot cavity (FPC) outside the
laser~\cite{SizerT_89a,UdemT_99a} was suggested as one
solution~\cite{MurphyM_07e,SchmidtP_08a} and has proven
effective~\cite{LiC-H_08a,BrajeD_08a}. The FPC comprises two mirrors
separated by a distance smaller than the laser cavity length so that
all modes, except every $m$th ($m>1$), are interferometrically
suppressed (Fig.~1, lower panel). We used a FPC stabilized to a filter
ratio, $m$, by controlling its length with an electronic servo system
to generate effective mode spacings, $m \times f_{\rm rep}$, between 1
and 15\,GHz. The degree to which the unwanted modes are suppressed is
an important parameter: The FPC transmission function falls sharply
away from the transmitted mode frequencies but, because nearby suppressed
modes are not resolved from the transmitted ones by the spectrograph,
small asymmetries in this function (especially combined with
time variations) can cause systematic shifts in the measured line
positions. With our setup, we achieve an unwanted mode suppression of
more than 46~dB at filter ratios $m \leq 20$.  Other possible
systematic shifts due to the filtering have been
identified~\cite{BrajeD_08a} and need to be controlled.

LFC spectra were recorded with and without the spectrum of a small
section of the Sun's photosphere at wavelengths $\sim$1.5\,$\upmu$m. A
sample $m \times f_{\rm rep}=15$-GHz recording, superimposed with
Fraunhofer and atmospheric lines, is shown in Fig.~2. To estimate our
calibration accuracy and to test the spectrograph's stability, we
analyzed several exposures of the LFC only. Individual Lorentzian
functions were fitted to the recorded modes as a function of pixel
position and identified with the absolute comb frequencies, $f_n$,
which are referenced to the atomic clock~\cite{materials}. The dense
grid of modes allows the spectrograph's calibration function (Fig.~3A) to be determined to very high accuracy; even a simple,
second-order polynomial fit to the pixel-versus-frequency distribution
has only 9\,m\,s$^{-1}$ root mean square (RMS) residual deviations
around it (Fig.~3B), and this remains almost unchanged with
higher-order polynomial modeling~\cite{materials}.

With traditional calibration techniques, such as thorium comparison
lamps, I$_2$ gas absorption cells or Earth's atmospheric
absorption lines for calibration achieve $\sim$10\,m\,s$^{-1}$
absolute precision per calibration line at best~\cite{LovisC_07a}.
Thus, even with these "first light" comb recordings, we already
demonstrate superior absolute calibration accuracy. Because more than $10^{4}$
modes will be available in a larger-bandwidth LFC, the statistical
uncertainty would be reduced to the 1\,cm\,s$^{-1}$ regime if the
residuals were truly random. However, the theoretical shot noise limit
calculated from the number of photons recorded per comb mode is much
smaller than 9\,m\,s$^{-1}$; systematic effects from the spectrograph
and detector system evidently completely dominate the residuals.

The main reason for testing LFCs at real telescopes, on real
astronomical spectrograph and detector systems, is to understand how
to measure and then mitigate and/or model and remove such systematics.
Because the VTT spectrograph is not stabilized (i.e.~temperature-,
pressure- and vibration-isolated), instrument drifts are expected and
the VTT LFC spectra can already be used to track them accurately.
From a time series of exposures, we derive a drift in the spectrograph
of typically 8\,m\,s$^{-1}$\,min$^{-1}$ (5\,MHz\,min$^{-1}$)~\cite{materials}. Much lower drift rates have been demonstrated with suitably stabilized instruments
[e.g.~$\sim$1\,m\,s$^{-1}$ over months with HARPS~\cite{LovisC_06a}];
although the VTT is not optimized for stability, this does not affect
its usefulness to test calibration procedures. Indeed, different modes
are observed to drift at different rates, with neighboring modes having
highly correlated drift rates~\cite{materials}. Also, as the comb
modes drift across the detector, higher-order distortions are evident,
which are the combined result of many effects, such as intrapixel
sensitivity variations. Thus, the VTT data already show an important
advantage of LFC calibration: The dense grid of high SNR calibration
information allows the discovery and measurement of complex effects correlated
across the chip and in time.

The first light for frequency combs on astronomical
spectrographs has delivered calibration precision beyond the state of
the art. The key opportunity now is to use LFC spectra to measure
and remove systematic effects in order to reach the
$\sim$1\,cm\,s$^{-1}$ long-term calibration precision, accuracy, and
repeatability required to realize the redshift drift experiment.


\bigskip

\noindent{\bf Supporting Online Material}\\
www.sciencemag.org/cgi/content/full/1161030/DC1\\
Materials and Methods\\
Figs.~S1 to S4\\
References\\
\\
28 May 2008; accepted 25 July 2008\\

\clearpage

\begin{figure}
\centerline{\includegraphics[width=0.95\columnwidth]{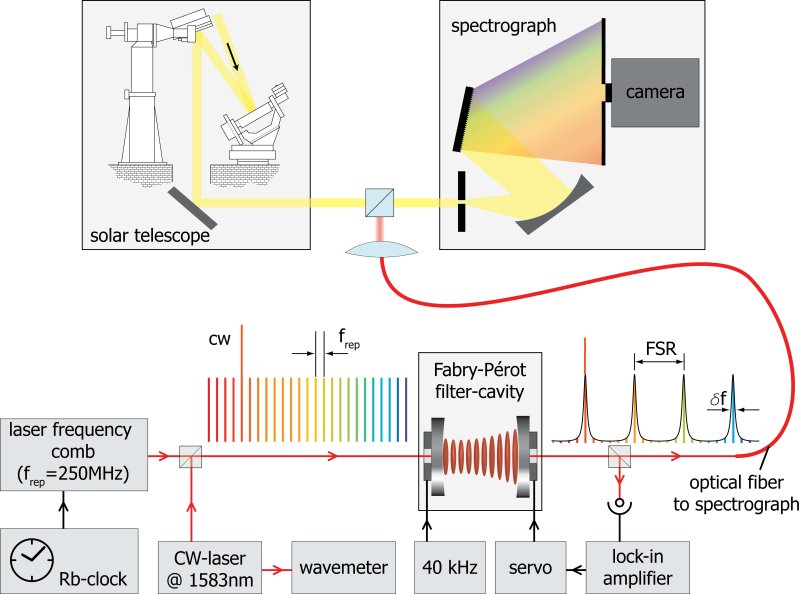}}
\caption{Sketch of our experimental setup at the VTT.  By
  superimposing the frequency comb with light from a celestial body --
  in this case, the Sun -- one can effectively calibrate its emission
  or absorption spectrum against an atomic clock. An erbium-doped
  fiber LFC with 250-MHz mode spacing (pulse repetition rate) is
  filtered with a FPC to increase the effective mode spacing, allowing
  it to be resolved by the spectrograph. The latter has a resolution
  of $\sim$0.8\,GHz at wavelengths around 1.5\,$\upmu$m, where our LFC
  tests were conducted. The LFC was controlled by a rubidium atomic
  clock. A continuous-wave (CW) laser at 1583\,nm was locked to one
  comb line and simultaneously fed to a wavemeter. Even though the
  wavemeter is orders of magnitude less precise than the LFC itself,
  it is sufficiently accurate (better than 250\,MHz) to identify the
  mode number, $n$. The FPC length, defining the final free spectral
  range (FSR), was controlled by feedback from its output.
  See~\cite{materials} for further details.}
\end{figure}

\begin{figure}
\centerline{\includegraphics[width=0.65\columnwidth]{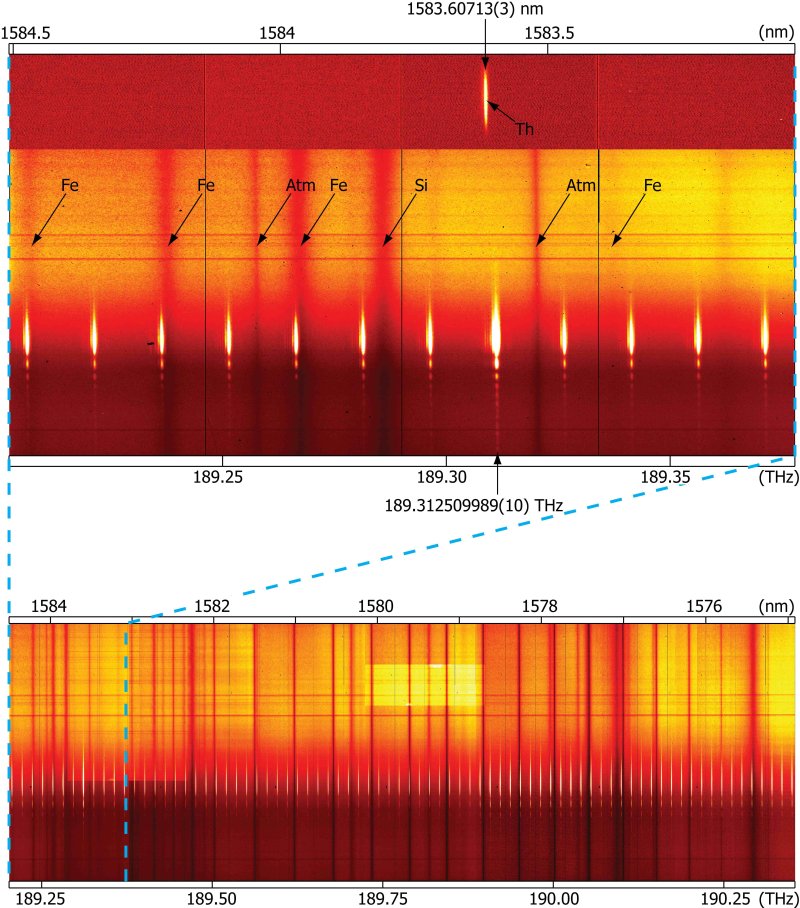}}
\caption{Spectra of the solar photosphere (background image) overlaid
  by a LFC with 15\,GHz mode spacing (white, equally spaced vertical
  stripes). Spectra are dispersed horizontally, whereas the vertical
  axis is a spatial cross section of the Sun's photosphere.  The upper
  panel shows a small section of the larger portion of the spectrum
  below. The brighter mode labeled with its absolute frequency is
  additionally superimposed with a CW laser used to identify the mode
  number (Fig.~1). The frequencies of the other modes are integer
  multiples of 15\,GHz higher (right) and lower (left) in frequency.
  Previous calibration methods would use the atmospheric absorption
  lines (dark vertical bands labeled ``Atm'' interleaved with the
  Fraunhofer absorption lines), which are comparably few and far
  between. Also shown in the upper panel is the only thorium emission
  line lying in this wavelength range from a typical hollow-cathode
  calibration lamp.  Recording it required an integration time of
  30\,min, compared with the LFC exposure time of just 10\,ms. Unlike
  with the LFC, the thorium calibration method cannot be conducted
  simultaneously with solar measurements at the VTT. The nominal
  horizontal scale is 1.5$\times$10$^{-3}$\,nm\,pixel$^{-1}$ with
  $\sim$1000 pixels shown horizontally in the upper panel. Black
  horizontal and vertical lines are artifacts of the detector array.}
\end{figure}

\begin{figure}
  \centerline{\includegraphics[width=0.55\columnwidth]{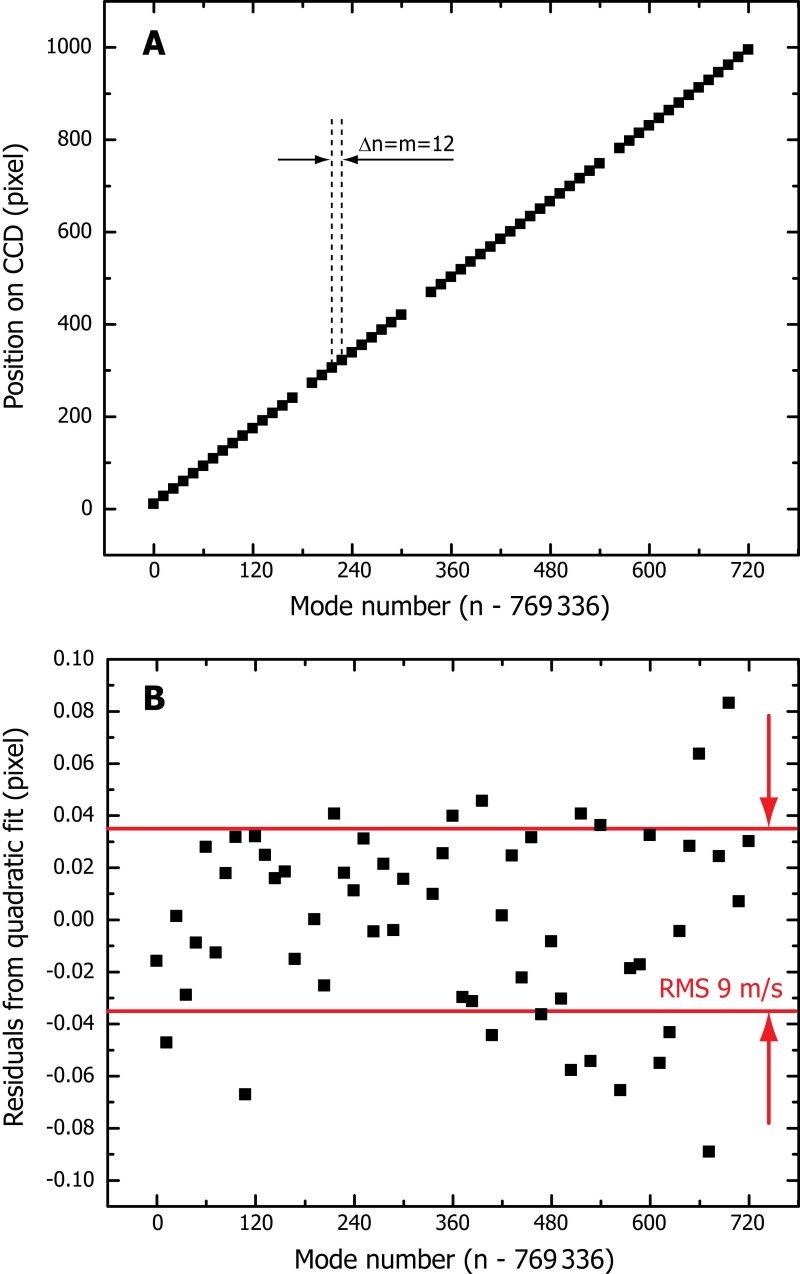}}
  \caption{Precision achieved with our calibration with a LFC filtered
    to 3\,GHz ($m=12$).(A) The position of the transmitted modes,
    derived from a multi-Lorentzian fit, plotted against the mode
    number.  Modes without a corresponding detector position
    measurement were deemed unsuitable for use in calibration because
    they fell on large detector artifacts and/or were overlaid with
    light from the CW laser.  The size of one pixel corresponds to
    172\,MHz at 1583\,nm. On this scale, no distortions are
    visible.(B) The residuals from a quadratic fit that gives a RMS
    residual of 9\,m\,s$^{-1}$.  The quadratic fit greatly reduces the
    residuals compared to a linear model, whereas higher-order
    polynomials do not improve the performance of the fit
    significantly~\cite{materials}. Even with these first LFC
    recordings from the VTT, the 9\,m\,s$^{-1}$ RMS residuals here
    indicate better absolute calibration than is achieved with
    traditional calibration methods~\cite{LovisC_07a}.}
\end{figure}

\clearpage

{\Huge\bf SUPPORTING ONLINE MATERIAL}

\section*{Materials and methods}

\subsection*{Laser frequency comb setup}

To use frequency combs for precise calibration of astronomical
spectrometers, the latter need to resolve apart the modes of the
former. Therefore, the comb's mode spacing must be larger than the
spectrometer's resolution. Most `high resolution' astronomical
spectrographs have resolving powers around $R\equiv\lambda/{\rm
  FWHM}\sim10^5$ (where FWHM is the spectrograph's
full-width-at-half-maximum wavelength resolution) to obtain the best
signal-to-noise ratio (SNR) for faint, distant astronomical objects.
In the infrared region around 1.5\,$\upmu$m, the required mode spacing
exceeds 2\,GHz while the optimum lies around $10$\,GHz for a typical
spectrometer operating in the visible~\cite{MurphyM_07e_S1}.

Unfortunately, a mode-locked laser with $f_{\rm rep}=10$\,GHz
corresponds to a laser cavity length of only 1.5\,cm. At present, all
lasers that reach such a large repetition rate are not useful for
calibrating astronomical spectrographs for other reasons: no $f_{\rm
  rep}=10$\,GHz LFC has yet reached the spectral bandwidth required to
provide a large set of lines and to self reference the frequency comb, i.e. to stabilize it to an atomic clock.
Current high-$f_{\rm rep}$ LFCs, such as Titanium--Sapphire LFCs, are
delicate devices that would not operate autonomously for extended
observation times at remote telescopes. The only lasers currently
complying with the latter requirement are mode locked fiber lasers.
But fiber LFCs with a wide spectral bandwidth have much lower
repetition rates than required and so constructing a high-$f_{\rm
  rep}$ LFC with octave-spanning bandwidth (for self-referencing) is a
major technical challenge~\cite{MurphyM_07e_S1}. So far, the highest
fundamental pulse repetition rate that has been demonstrated with
fiber lasers is 250\,MHz~\cite{WilkenT_07a_S2}.

Our approach to solve this problem is to filter the modes with a
Fabry-P\'{e}rot cavity (FPC) outside the LFC~\cite{MurphyM_07e_S1,SizerT_89a_S3,UdemT_99a_S4,SchmidtP_08a_S5,LiC-H_08a_S6}. Such
an interferometric cavity consists of two mirrors with reflectivity
$R$ separated by a distance $L$ that has a spectral transmission
function
\begin{equation}
\renewcommand \theequation{S1}
  T(f) = \frac{(1-R)^2}{(1-R)^2+4R\sin^2(2 \pi f L/c)}.
\end{equation}
This implies a regular grid of transmissions, each $\Delta f=FSR
(1-R)/\pi$ wide, with free spectral range $FSR=c/2L$. For our purpose
the filter mode spacing is set to an integer multiple $m$ of the
laser comb spacing by adjusting $L$, such that $m f_{\rm rep}=c/2L$.
The filter cavity then transmits exactly every $m$th mode while the
unwanted modes in between are largely suppressed.

Our $f_{\rm rep}=250$\,MHz, octave-spanning, erbium-doped fiber LFC
was self-referenced with the common $f$--$2f$
technique~\cite{UdemT_02a_S7}. A portable rubidium atomic clock with a
rather moderate accuracy of 5 parts in $10^{11}$ was used in this
demonstration to stabilize the repetition rate $f_{\rm rep}$ and the
carrier envelope offset frequency $f_{\rm ceo}$ of the comb. Much
better clocks are available if needed.  Stabilization of the
Fabry-P\'{e}rot cavity (FPC) onto the LFC was achieved by modulating
the FPC length at 40\,kHz using a piezo actuator onto which one of the
FPC mirrors is mounted. Part of the transmitted light is sent to a
photo-detector whose output is de-modulated with a lock-in amplifier
referenced to the modulation frequency. This generates a bi-polar
error signal which, using a second piezo actuator, keeps the FPC on
resonance. After this locking is achieved, the mode number needs to be
identified. This is complicated by the fact that the mode filter
cavity can fix onto any comb mode. Thus, a resolution of 250\,MHz
(i.e.~$f_{\rm rep}$) is required even though the final effective mode
spacing is larger. This is achieved by locking a continuous wave (CW)
laser to a comb line while its frequency is measured by a wavemeter.
Even though the wavemeter is orders of magnitude less precise than the
LFC itself, it is sufficiently accurate (better than 250\,MHz) to
identify the mode number.

As stated in the main text, it is important to suppress the unwanted
modes by a large factor so that they do not contribute to systematic
errors in the transmitted mode centroids as measured on the
astronomical detector. Fig.~S1 shows the spectral envelope of our mode
locked laser together with various filter settings, $m$. For large
filter ratios the mode spacing can even be resolved with a common
laboratory optical spectrum analyzer. Equation S1 states that the
laser modes closest to the transmitted mode will have the smallest
suppression. For this reason the direct optical spectrum does not
provide a good way to estimate the unwanted mode suppression ratio.
We used two experimental methods to obtain upper limits on this ratio.
Firstly, for direct detection of the unwanted modes in the optical
domain, we recorded heterodyne beat signals of the filtered LFC with a
CW laser. The unwanted modes are below our detection sensitivity which
is 30~dB as shown in Fig.~S2~(panel b).  Secondly, detecting the pulse
repetition rate of the filtered LFC allows to derive the suppression ratio in a rather
indirect way. Fig.~S2~(panel a) shows the radio frequency spectrum of
the filtered pulse train. The filtered mode spacing, $m\times f_{\rm
  rep}$, can be understood as being due to the beat frequency between
these modes. The component at $f_{\rm rep}$ in the spectrum derives
mainly from the beating between the transmitted modes and the
suppressed modes closest to them. With everything else ideal the
observed mode suppression, as measured in the radio frequency
spectrum, appears 6\,dB less efficient than for the true optical mode
suppression~\cite{ChenJ_08a_S8}.

In future, there might be other useful approaches to reach a
sufficient mode spacing aside from using FPCs. Higher repetition rate
lasers and fiber lasers maintaining a string (rather than a single)
intra-cavity pulse exist, but are currently difficult to employ for
astronomical calibration as they produce only a small spectral
bandwidth or lack effective methods for suppressing extraneous modes.
Another promising technique, albeit in its technological infancy,
relies on parametric frequency conversion microresonators~\cite{DelHayeP_07a_S9}.

\subsection*{The German Vacuum Tower Telescope}

The VTT~\cite{SchroeterE_85a_S10} has a 15\,m vertical spectrograph which,
with a IR-optimized diffraction grating working in fifth order,
provides a very high spectral resolution, $R\approx3\times10^5$. For
1.5\,$\upmu$m observations, a TCM8000, 1K$\times$1K infrared array from
Rockwell Scientific is used as a detector. With an 18.5\,$\upmu$m square
pixel size it provides a nominal scale of 1.5$\times
10^{-3}$\,nm\,pixel$^{-1}$. The spectrograph is pre-dispersed, so only
a fraction of one echelle order is recorded simultaneously, covering
about 1.5\,nm. The filtered LFC light was superimposed with the Sun
light by means of a beam splitter cube $\sim$1\,m in front of the
spectrograph's 40\,$\upmu$m entrance slit.

\subsection*{VTT calibration and systematic errors with LFC spectra}

To demonstrate how well the VTT spectrograph's wavelength scale could
be defined using the LFC, we performed a wavelength calibration
procedure on a $m\times f_{\rm rep}=3$\,GHz recording similar to that
carried out with typical modern astronomical long-slit spectra. The
sum of the pixel photon-counts in 100 rows of the detector was used to
extract a 1-dimensional comb spectrum.  Lorentzian functions (with
width $\sim$3 pixels) were fitted to the individual comb modes to
define their centroids. A similar extraction and fit is shown in the
upper panel of Fig.~S3 for a $m\times f_{\rm rep}=15$\,GHz recording.
Note that the crude extraction technique picks out flaws in the
detector and extraneous scattered light; no effort was made to remove
these by traditional techniques (e.g.~flat-fielding) but obviously
spurious data was not included in the multi-Lorentzian fit.  Again,
for demonstration purposes, this is adequate. The fitted centroids (in
pixels) of the comb modes were then plotted against their known
frequencies in Hz to form the calibration function of the recording.
Polynomial fits of varying order were fitted to the calibration
function but it was found that the residuals around the fit reduced
only marginally for orders higher than 2 -- see Fig.~S4.  Even around
the second order fit, the calibration residuals had a RMS of only
9\,m\,s$^{-1}$.

As described in the main text, the VTT data are also useful to
demonstrate a key advantage of LFC calibration: the mapping out of
systematic effects in astronomical spectrographs and detector systems.
Since the VTT is not optimized for stability, the largest systematic
effect was a small drift of the comb lines across the detector array.
The drift was best measured in a series of recordings made with
$m\times f_{\rm rep}=15$\,GHz as shown in Fig.~S3. Even though the
drift was only $\sim$0.2\,pixels overall, it was easily detected with
high precision due to the large amount of calibration information
afforded by the LFC modes. Moreover, as can be seen in Fig.~S3, there
is a distortion in the drift as a function of position along the
detector; the drift rate is higher for some parts of the chip. This
effect is evident only because it is seen to be correlated among
neighboring comb modes. That is, again, the dense calibration
information of the LFC allows systematics to be detected.

Of course, with a spectrograph and detector system optimized for
stability, the aim would be to largely suppress such effects and
precisely measure the residuals. Nevertheless, systematic errors
similar to these are still likely to be present at some level larger
than 1\,cm\,s$^{-1}$. For example, inhomogeneities in the
spectrograph's gratings, particularly their rulings, may produce
small-scale distortions in the spectrograph's calibration curve.
Intra-pixel sensitivity variations will also be important, as will
variations in the pixel size and/or spacing between pixels. By making
small adjustments in either $f_{\rm rep}$ and/or $f_{\rm ceo}$ in a
series of `off-line' calibration exposures (i.e.~not taken
simultaneously with the science object exposure), a complete and
detailed `map' of these effects across the chip can be generated. More
difficult to measure and characterize will be time-variations in these distortions.
To this end, frequency comb light should be recorded simultaneously
with the science object light on the same detector.
Of course, the distortions affecting the science spectrum will be very
slightly different to those affecting the comb spectrum because of the
necessary (but small) spatial separation between them on the detector.
These can be corrected to first order using the map above; trend
analysis of off-line time-series calibrations should allow higher
order corrections.

Comb modes with a finite width are succeptable to yet another
systematic effect if correlated frequency and amplitude modulation is
present. In this case the spectral center of mass used for calibration
could be shifted relative to the average frequency that is referenced
to the atomic clolck.  Non perfect alignmet of the FPC modes to the
comb modes can be the cause of such a correlated modulation as pointed
out by D.~A.~Braje~et~al.~\cite{BrajeD_08a_S11}. We believe that this
effect can be reduced to insignificant levels, for example by
narrowing the linewidths of the comb modes.

Still, there may be systematic errors which are difficult to map out
(spatially and temporally) with the comb light. For example, one might
imagine that some errors occur during the read-out of the detector.
It is also likely that some important systematic errors are yet to be
discovered or conceived yet. This further illustrates the importance
of beginning detailed analyses of frequency comb spectra taken on
astronomical detectors in order to test the real precision level
achievable before the final designs (and therefore key science
activities) of the future generation of telescopes have been
determined. Even these first light VTT recordings demonstrate how LFC
calibration can be used not only to define the wavelength scale but
also as a tool for discovering systematic effects.  Only by future,
more extensive investigations of this nature can we know whether
removing systematics at the 1\,cm\,s$^{-1}$ precision level is really
achievable.

\clearpage

\begin{figure}
\renewcommand \thefigure{S1}
\centerline{\includegraphics[width=0.75\columnwidth]{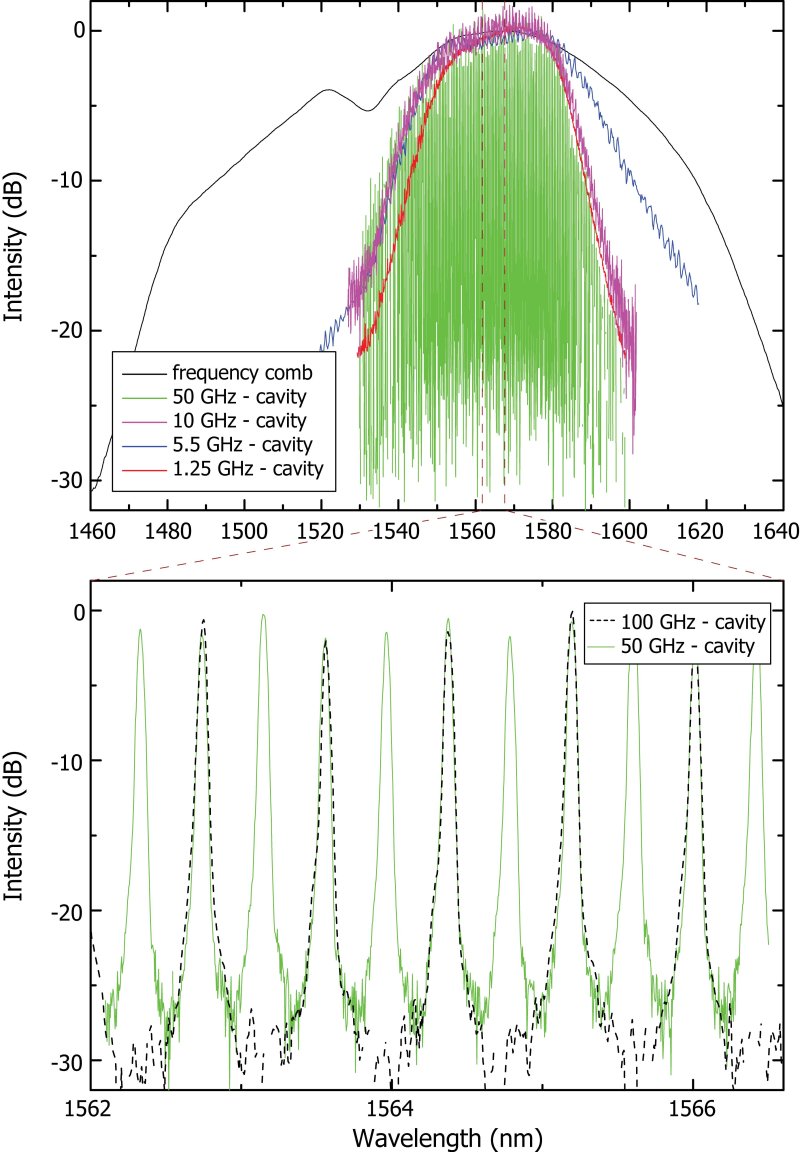}}
\caption{The upper panel shows a spectrum of the
optical frequency comb as emitted by the laser (black line) and the
transmission spectra of the Fabry-P\'{e}rot cavity (colored lines)
made from two $R = 99.87\%$ mirrors.  Various mode filter
ratios, $m = {5, 22, 40}$ and\,$200$, are investigated but the
individual modes are resolved by the optical spectrum analyzer only
for filter ratios $m \geq 200$. The lower panel shows a small section
of the full spectrum above. For filter ratios $m = {200}$\,and\,$400$,
corresponding to filtered mode spacings of 50\,GHz and 100\,GHz
respectively, a full modulation of the optical spectrum is clearly
visible.}
\end{figure}

\begin{figure}
\renewcommand \thefigure{S2}
\centerline{\includegraphics[width=0.95\columnwidth]{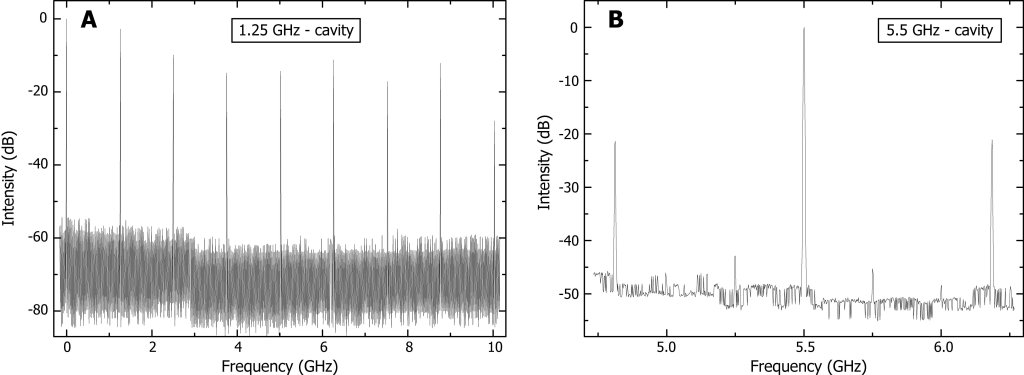}}
\caption{Panel (a) shows the radio frequency spectrum
of a 1.25\,GHz ($m = 5$) filtered pulse train. Many harmonics of
$m\times f_{\rm rep}$ are seen but the unwanted modes at 250\,MHz are
below the detection limit of 60\,dB. In panel (b) we have changed the
filter ratio to $m = 22$ and added a continuous wave (CW) laser for
heterodyning.  One can see the 5.5\,GHz mode spacing of the filtered
comb as well as the unwanted modes at 5.25\,GHz and 5.75\,GHz
suppressed by 43\,dB. The heterodyne signal at around 6.4\,GHz and
4.6\,GHz is due to the beating of the CW laser with the main
transmitted modes of the filtered LFC. The beating with the unwanted
modes is below the detection limit, i.e. 30\,dB in this case.}
\end{figure}

\begin{figure}
\renewcommand \thefigure{S3}
\centerline{\includegraphics[width=0.95\columnwidth]{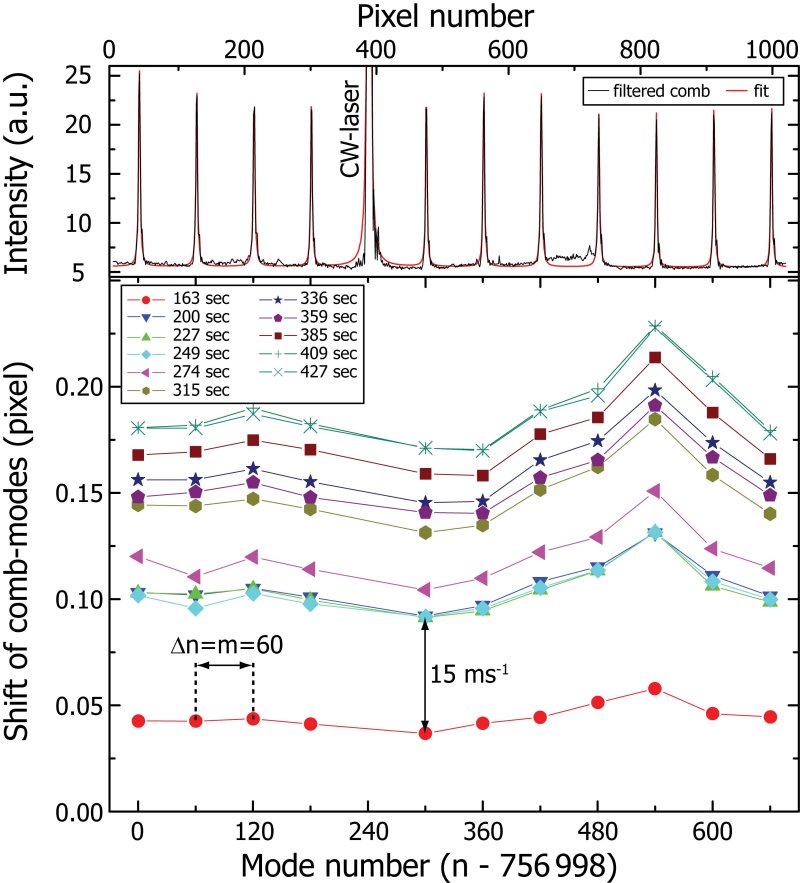}}
\caption{The upper panel shows the VTT spectrum of the
LFC filtered to 15\,GHz ($m = 60$) where the signal has been integrated along
over 100 rows of the detector. A sum of Lorentzian functions has been
fitted to the data to determine the line center positions. The distance
between adjacent pixels corresponds to 172\,MHz at 1583\,nm. The lower panel shows
the fitting results from a time series of these calibration
measurements; the line center positions are shown with respect to a
reference measurement. The drift is distorted along the detector (some
modes drift faster than others) and varies non-linearly with time. An
average drift of 8\,m\,s$^{-1}$\,min$^{-1}$ was derived from this time
series.}
\end{figure}

\begin{figure}
\renewcommand \thefigure{S4}
\centerline{\includegraphics[width=0.95\columnwidth]{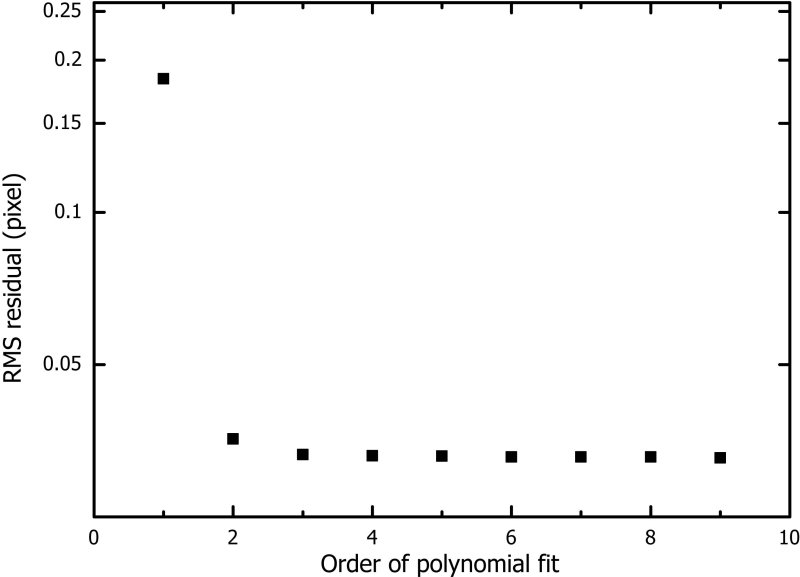}}
\caption{Polynomials of different orders were fitted
to the calibration function of the detector as measured with a comb
spacing of 3\,GHz (i.e.~58 modes were included in the fits). The pixel
size in this configuration corresponds to 187\,MHz at 1558\,nm. The
RMS residual is plotted against the order of the polynomial. A
quadratic fit to the calibration function is already a good
approximation and provides RMS residuals of 9\,m\,s$^{-1}$.}
\end{figure}

\end{document}